\documentclass[times,twocolumn,final]{elsarticle}

\usepackage{medima}
\usepackage{framed,multirow}

\usepackage{amssymb}
\usepackage{amsmath}
\usepackage{latexsym}
\usepackage{algorithm}
\usepackage{algpseudocode}
\usepackage{dblfloatfix}

\usepackage{array}
\newcolumntype{M}[1]{>{\centering\arraybackslash}m{#1}}

\usepackage{url}
\usepackage{xcolor}

\usepackage{hyperref}

\definecolor{newcolor}{rgb}{.8,.349,.1}

\journal{Medical Image Analysis}

\begin{document}

\verso{Siyuan Dong \textit{et~al.}}

\begin{frontmatter}

\title{A Flow-based Truncated Denoising Diffusion Model for Super-resolution Magnetic Resonance Spectroscopic Imaging\tnoteref{tnote1}}%

\author[1]{Siyuan \snm{Dong}\corref{cor1}}
\cortext[cor1]{Corresponding author at: 300 Cedar Street, New Haven, CT, 06519}
\ead{s.dong@yale.edu}

\author[2,9]{Zhuotong \snm{Cai}\corref{cor1}}
\ead{zhuotong.cai@yale.edu}
\author[3,4]{Gilbert \snm{Hangel}}
\author[3]{Wolfgang \snm{Bogner}}
\author[4]{Georg \snm{Widhalm}}
\author[5]{Yaqing \snm{Huang}}
\author[2]{Qinghao \snm{Liang}}
\author[1]{Chenyu \snm{You}}
\author[6]{Chathura \snm{Kumaragamage}}
\author[6]{Robert K. \snm{Fulbright}}
\author[6]{Amit \snm{Mahajan}}
\author[1,7,8]{Amin \snm{Karbasi}}
\author[6]{John A. \snm{Onofrey}}
\author[2,6]{Robin A. \snm{de Graaf}}
\author[1,2,6]{James S. \snm{Duncan}\corref{cor1}}
\ead{james.duncan@yale.edu}

\address[1]{Department of Electrical Engineering, Yale University, New Haven, CT, USA}
\address[2]{Department of Biomedical Engineering, Yale University, New Haven, CT, USA}
\address[3]{Department of Biomedical Imaging and Image-Guided Therapy, Highfield MR Center, Medical University of Vienna, Vienna, Austria}
\address[4]{Department of Neurosurgery, Medical University of Vienna, Vienna, Austria}
\address[5]{Department of Pathology, Yale University, New Haven, CT, USA}
\address[6]{Department of Radiology and Biomedical Imaging, Yale University, New Haven, CT, USA}
\address[7]{Department of Computer Science, Yale University, New Haven, CT, USA}
\address[8]{Department of Statistics and Data Science, Yale University, New Haven, CT, USA}
\address[9]{Institute of Artificial Intelligence and Robotics, Xi’an Jiaotong University, Xi'an, China}

\received{xxx}
\finalform{xxx}
\accepted{xxx}
\availableonline{xxx}
\communicated{xxx}

\begin{abstract}
Magnetic Resonance Spectroscopic Imaging (MRSI) is a non-invasive imaging technique for studying metabolism and has become a crucial tool for understanding neurological diseases, cancers and diabetes. High spatial resolution MRSI is needed to characterize lesions, but in practice MRSI is acquired at low resolution due to time and sensitivity restrictions caused by the low metabolite concentrations. Therefore, there is an imperative need for a post-processing approach to generate high-resolution MRSI from low-resolution data that can be acquired fast and with high sensitivity. Deep learning-based super-resolution methods provided promising results for improving the spatial resolution of MRSI, but they still have limited capability to generate accurate and high-quality images. Recently, diffusion models have demonstrated superior learning capability than other generative models in various tasks, but sampling from diffusion models requires iterating through a large number of diffusion steps, which is time-consuming. This work introduces a Flow-based Truncated Denoising Diffusion Model (FTDDM) for super-resolution MRSI, which shortens the diffusion process by truncating the diffusion chain, and the truncated steps are estimated using a normalizing flow-based network. The network is conditioned on upscaling factors to enable multi-scale super-resolution. To train and evaluate the deep learning models, we developed a \textsuperscript{1}H-MRSI dataset acquired from 25 high-grade glioma patients. We demonstrate that FTDDM outperforms existing generative models while speeding up the sampling process by over 9-fold compared to the baseline diffusion model. Neuroradiologists' evaluations confirmed the clinical advantages of our method, which also supports uncertainty estimation and sharpness adjustment, extending its potential clinical applications.
\end{abstract}


\end{frontmatter}


\section{Introduction}
Magnetic Resonance Spectroscopic Imaging (MRSI) is a non-invasive imaging technique for studying metabolism within the body. Since the study of metabolism plays a key role in understanding a range of diseases including neurodegenerative diseases, cancers and diabetes, MRSI has become a valuable clinical and preclinical tool. MRSI is feasible on any nucleus possessing nuclear spin, and the most commonly used nucleus for in vivo MRSI is proton (\textsuperscript{1}H) due to its high nuclear magnetic resonance (NMR) sensitivity and abundance in metabolites \citep{de2019vivo}.

High spatial resolution MRSI is needed to characterize small lesions and intra-lesion heterogeneity. However, in practice, MRSI is acquired at low resolution due to time and sensitivity restrictions caused by the low metabolite concentrations. To achieve a satisfactory signal-to-noise ratio (SNR) with clinical practice, which often employs 3T scanners and mandates acceptable scan times, the spatial resolution of \textsuperscript{1}H-MRSI is often restricted to approximately 1$\times$1$\times$1 cm$^3$. While advancements in hardware and acceleration techniques have been noteworthy \citep{bogner2021accelerated}, achieving high-resolution MRSI through stronger magnetic fields (e.g., 7T) or extended scan times remains largely impractical for routine clinical use, especially on established 3T MRI systems. Therefore, developing a post-processing approach to generate high-resolution MRSI from low-resolution scans will greatly benefit its clinical applications.

Traditional post-processing approaches for super-resolution MRSI primarily employ model-based regularization, which incorporates high-resolution anatomical information from corresponding MRI scans and enforces spatial smoothness using total variation distance \citep{lam2014subspace,jain2017patch,kasten2016magnetic,hangel2019high}. However, there are a few problems with these regularization-based methods: (1) While a correlation exists between MRI intensity and MRSI metabolite concentrations, MRIs lack metabolic information. For this reason, over-reliance on MRI-based regularization might bias the resulting high-resolution MRSI towards the MRI \citep{dong2023preserved}. (2) Handcrafted regularizations, such as total variation distance, do not truthfully reflect the intrinsic characteristics of high-resolution MRSI, potentially leading to over-smoothed results. (3) Each new acquisition needs to undergo an optimization process, which is time-consuming.

Deep learning has shown great potential in super-resolution tasks for medical imaging \citep{li2021review} and can effectively circumvent the limitations of conventional techniques. The first deep learning method for super-resolution MRSI \citep{iqbal2019super} utilized a Dense UNet to map low-resolution MRSI metabolic maps to higher resolution ones by incorporating anatomical information from T1 MRI. However, due to the absence of a public MRSI training dataset, these metabolic maps were synthesized from MRI, which has a fundamental issue that the simulated MRSI may not authentically capture all metabolite characteristics. To address this issue, \citet{dong2021high} introduced an in vivo \textsuperscript{1}H-MRSI dataset acquired from high-grade glioma patients to train a super-resolution MRSI network. This work aimed to restore high-resolution details and enhance visual quality by applying adversarial loss, which was also explored in subsequent research \citep{dong2022multi,li2022deep}. Imposing adversarial loss \citep{wang2020deep,chen2024mri} via Generative Adversarial Networks (GAN) \citep{goodfellow2020generative} is often subject to training instability and mode-collapse. To more accurately learn the distribution of high-resolution MRSI images, \citet{dong2022flow} proposed a normalizing flow-based network. This framework learns the target distribution via likelihood maximization, which makes the training more stable and interpretable, outperforming GAN-based approaches. Nonetheless, the flow-based methods must be implemented by invertible neural networks \citep{ardizzone2018analyzing,lugmayr2020srflow,liang2021hierarchical,dong2022invertible,chen2024systems}, which has constrained architecture and limit the learning capability.

Recently, diffusion models \citep{sohl2015deep,ho2020denoising,song2019generative,song2020score} have demonstrated superior learning capability than other generative models in a range of tasks \citep{dhariwal2021diffusion,yang2022diffusion}. Diffusion models can be used to learn conditional image distributions for image-to-image translation tasks, such as image super-resolution \citep{saharia2022image,li2022srdiff} and MRI reconstruction \citep{chung2022score,peng2022towards,kazerouni2022diffusion}. A popular subclass of diffusion models is called Denoising Diffusion Probabilistic Model (DDPM) \citep{ho2020denoising,nichol2021improved}, which involves a forward diffusion process and a reverse diffusion process. The forward diffusion process gradually perturbs the target images with Gaussian noise through a large number of diffusion steps until a pure Gaussian noise is obtained. During the reverse diffusion process, a neural network retraces this process, training to revert the Gaussian noise back to the original image by gradually eliminating the noise. However, an inherent problem with diffusion models is that the sampling process, or the reverse diffusion process, requires iterating through a large number of diffusion steps, typically in the thousands, which is time-consuming. This lengthy process is due to the need to inject only a small amount of noise in each step to ensure that the forward and reverse processes have approximately the same functional form \citep{ho2020denoising}. One simple way to expedite the sampling process is to use strided steps \citep{nichol2021improved}, but this may compromise the quality of the samples. Recently, \citet{zheng2022truncated} proposed a Truncated Diffusion Probabilistic Model (TDPM) to shorten the diffusion process by truncating the diffusion chain, which achieved a similar level of performance as DDPM while achieving a significant speedup.

\begin{figure}
\centering
\includegraphics[width=8.7cm]{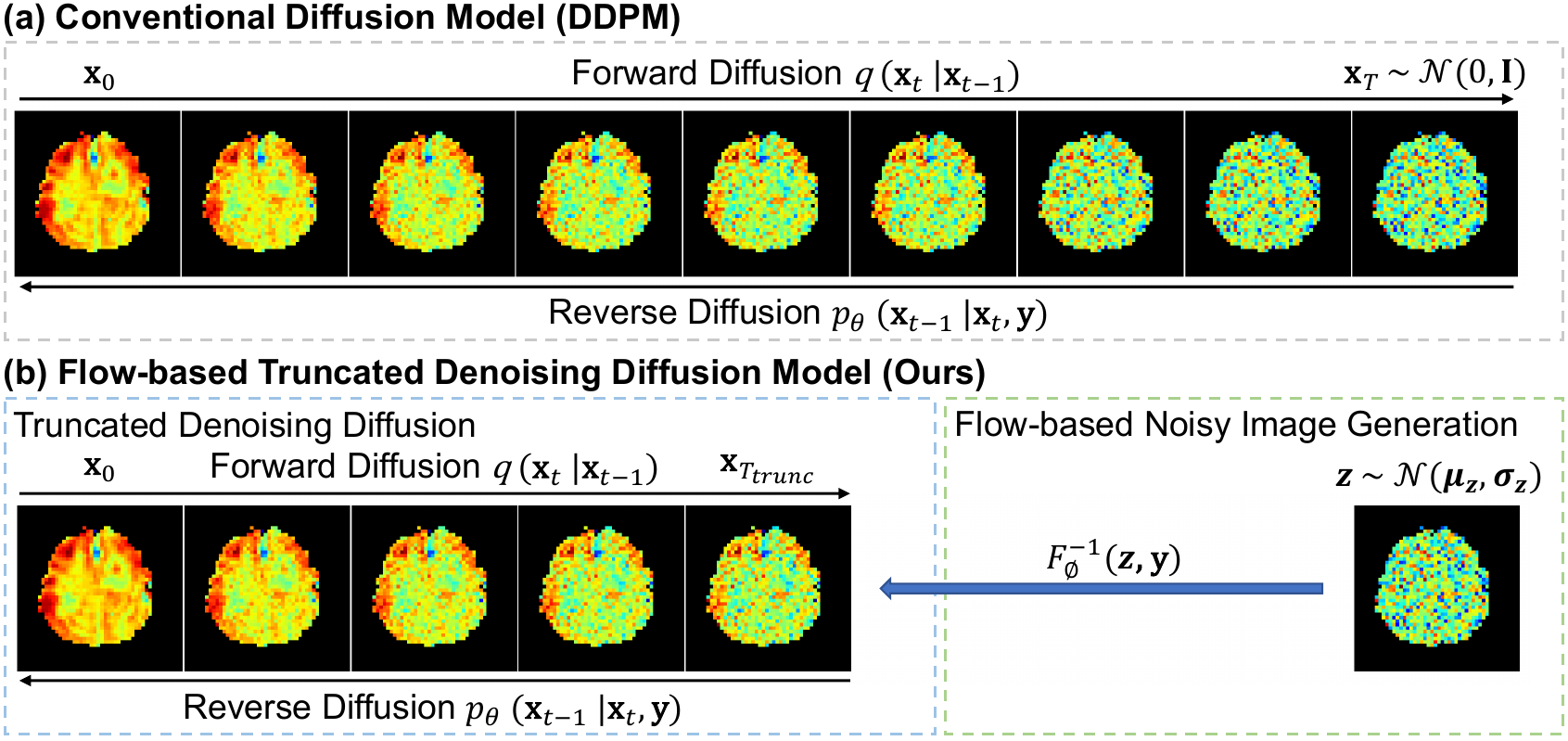}
\caption{A comparison between (a) the conventional diffusion model DDPM and (b) our method FTDDM. $\mathbf{x}_0$ is the noiseless high-resolution MRSI metabolic map. The forward diffusion process gradually adds Gaussian noise to $\mathbf{x}_0$. The noise is only added to the region of interest, as defined by a quality-filtering mask, to avoid the necessity of suppressing noise from the background during the reverse diffusion process. The reverse diffusion process uses a denoising network with parameters $\theta$ to retrace the forward diffusion process, provided with any condition images $\mathbf{y}$. $F^{-1}_{\phi}$ is the inverse of a normalizing flow-based network used to bridge the gap between the pure Gaussian noise $\mathbf{z}$ and the noisy image at the truncation point $\mathbf{x}_{T_{trunc}}$.}
\label{fig1}
\end{figure}

In this paper, inspired by TDPM, we propose a Flow-based Truncated Denoising Diffusion Model (FTDDM) for super-resolution MRSI. This model shortens the diffusion process of DDPM by truncating the diffusion chain, and the truncated steps are estimated using a normalizing flow-based network, as depicted in Fig.~\ref{fig1}. In contrast to the TDPM, which uses an adversarial network, our approach leverages a normalizing flow-based network to bridge the gap between pure Gaussian noise and the noisy image at the truncation point. This is because flow-based networks were trained with likelihood maximization, which is more stable than adversarial training, and demonstrated superior performance over GAN in the recent MRSI super-resolution work \citep{dong2022flow}. In addition, we condition our network on the upscaling factor using Conditional Instance Normalization. This promotes multi-scale super-resolution, eliminating the need to train a separate network for each upscaling factor. Moreover, we incorporate a temperature parameter to flexibly adjust the sharpness of the output images and investigate the trade-off between image visual quality and fidelity. To train and evaluate the proposed method, we develop a \textsuperscript{1}H-MRSI dataset acquired from 25 high-grade glioma patients. Experimental results show that FTDDM outperforms existing generative models, i.e. GAN and flow, while accelerating the sampling process by over 9-fold relative to DDPM. Neuroradiologists were invited to evaluate our method from a range of clinical perspectives, which confirms the clinical advantages of our super-resolved MRSI.

Our most significant contributions are: (1) development of a novel Flow-based Truncated Denoising Diffusion Model (FTDDM) for performant and sampling-efficient super-resolution MRSI; (2) adoption of network conditioning to facilitate multi-scale super-resolution; (3) introduction of a temperature parameter for sharpness adjustment; (4) development of an in vivo \textsuperscript{1}H-MRSI dataset acquired from glioma patients for training deep learning networks; and (5) Integration of neuroradiologists' ratings to assess the quality of super-resolution metabolic maps relative to the ground truth from clinical perspectives.

\section{Material and methods}

We first describe the acquisition and pre-processing of our dataset in Section \ref{sec_dataset}. We then present our Truncated Denoising Diffusion model and the corresponding loss function in Section \ref{sec_TruncDenoising}, followed by how mutli-scale super-resolution is incorporated into the network in Section \ref{sec_multiscale}. Section \ref{sec_flow} describes the design and training of the flow-based model for generating noisy images at the truncation point from Gaussian noise. Finally, the overall algorithm to sample from FTDDM is summarized in Section \ref{sec_sampling}.

\begin{figure*}
\centering
\includegraphics[width=\textwidth]{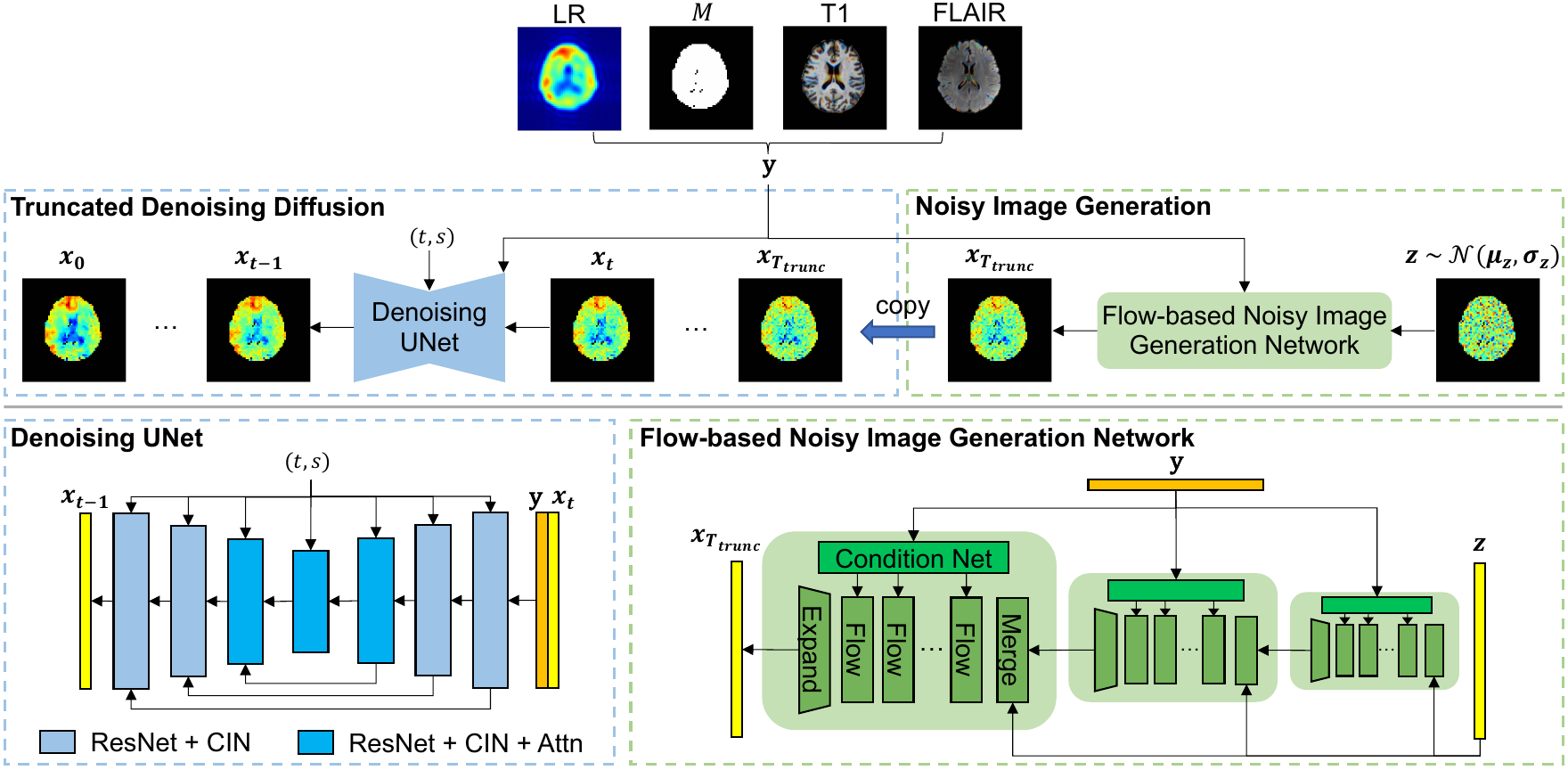}
\caption{Overview of the proposed method. The truncated denoising diffusion employs a Denoising UNet to iteratively estimate and remove noise from $\mathbf{x}_{T_{trunc}}$, resulting in a noiseless high-resolution MRSI metabolic map $\mathbf{x}_0$. The Denoising UNet also takes the condition $\mathbf{y}$, which is a concatenation of the low-resolution (LR) metabolic map, a quality-filtering mask $M$, T1 MRI and FLAIR MRI, i.e. $\mathbf{y}=\{$LR, $M$, T1, FLAIR$\}$. The Denoising UNet consists of Residual Network (ResNet) blocks and Conditional Instance Normalization (CIN). The CIN embeds timestep $t$ and upscaling factor $s$ into the network. The blocks in the middle have multi-head attention (Attn) modules, following \citet{nichol2021improved}. $\mathbf{x}_{T_{trunc}}$ is generated from the Gaussian noise $\mathbf{z}$ via the flow-based noisy image generation network, which comprises a series of flow layers across multiple dimensions, in line with \citet{dong2022flow}. Each flow layer contains conditional affine coupling, affine injector, invertible 1 × 1 convolution and activation normalization \citep{lugmayr2020srflow}. The condition images $\mathbf{y}$ are infused into the flow layers through Condition Networks (Condition Net), which consist of convolution layers and LeakyReLU.}
\label{fig2}
\end{figure*}

\subsection{Dataset}
\label{sec_dataset}
3D high-resolution \textsuperscript{1}H-MRSI, T1-weighted (MP2RAGE) and Fluid Attenuated Inversion Recovery (FLAIR) scans from 25 high-grade glioma patients were acquired on a 7T whole body MRI (Magnetom, Siemens Healthcare) equipped with a 32-channel receive coil array (Nova Medical). IRB approval and informed consent from all participants were obtained. MRSI sequence used spatial-spectral encoding to acquire 3D metabolic maps with a 64×64×39 measurement matrix and 3.4×3.4×3.4 mm³ nominal resolution in 15 minutes \citep{hangel2020high,hingerl2020clinical,dong2021high}. This resolution is notably high for MRSI due to the low SNR of metabolite signals. Additional parameters included an acquisition delay of 1.3 ms, a repetition time (TR) of 450 ms and a spectral bandwidth of 2778 Hz. 

T1 and FLAIR MRIs were skull-stripped and co-registered to MRSI using FSL v5.0 \citep{smith2004advances}. The processed MRSI spectra, as described in \citet{hangel2020high,hingerl2020clinical}, were quantified using LCModel v6.3-1 \citep{provencher2014lcmodel} to obtain 3D metabolic maps. For each of these maps, voxels with insufficient quality (SNR $<$ 2.5, or FWHM $>$ 0.15 ppm due to B$_0$ field inhomogeneity, or with severe extracranial lipid contamination proximal to the skull) were filtered out using a quality-filtering mask. In this work, we focus on 7 metabolites considered as major markers of onco-metabolism and proliferation \citep{hangel2020high}: total creatine (tCr), total choline (tCho), N-acetylaspartate (NAA), glutamate (Glu), inositol (Ins), glutamine (Gln), and glycine (Gly).

From each 3D MRSI scan of the 25 patients, we selected 9$\sim$18 axial slices with clear spatial characteristics of brain. Each slice provided 7 metabolites, resulting in a total of 2275 2D metabolic maps. These 64×64 metabolic maps are regarded as high-resolution ground truth, from which we derived low-resolution images using k-space truncation. The size of the truncation window depends on the upscaling factor, which will be discussed later.

\subsection{Truncated Denoising Diffusion}
\label{sec_TruncDenoising}
The conventional diffusion model, DDPM \citep{ho2020denoising}, adopts a forward and a reverse diffusion process to establish a mapping between a target distribution and a simple Gaussian distribution, as illustrated in Fig.\ref{fig1}(a). Starting with a high-resolution metabolic map $\mathbf{x}_0 \sim q\left(\mathbf{x}_0\right)$, the forward diffusion process gradually adds Gaussian noise to $\mathbf{x}_0$ over $T$ diffusion steps until a pure Gaussian noise $\mathbf{x}_T$ is achieved
\begin{equation}
\begin{aligned}
q\left(\mathbf{x}_{1: T} \mid \mathbf{x}_0\right) & =\prod_{t=1}^T q\left(\mathbf{x}_t \mid \mathbf{x}_{t-1}\right), \\
q\left(\mathbf{x}_t \mid \mathbf{x}_{t-1}\right) & =\mathcal{N}\left(\mathbf{x}_t \mid \sqrt{1-\beta_t} \mathbf{x}_{t-1}, \beta_t \mathbf{I}\right)
\end{aligned}
\end{equation}
where $\beta_t \in (0, 1)$ is the predefined noise variance at each timestep $t$. Based on the property of Gaussian distribution, we can obtain an analytical form of the forward process for an arbitrary timestep $t$
\begin{equation}
\label{eqn2}
q\left(\mathbf{x}_t \mid \mathbf{x}_{0}\right) =\mathcal{N}\left(\mathbf{x}_t \mid \sqrt{\gamma_t} \mathbf{x}_{0}, (1-\gamma_t) \mathbf{I}\right)
\end{equation}
where $\gamma_t=\prod_{i=1}^T (1-\beta_i)$. The reverse diffusion process counteracts the forward diffusion process by gradually eliminating noise from $\mathbf{x}_{T} \sim \mathcal{N}(0, \mathbf{I})$ to ultimately approximate $\mathbf{x}_0$, given any condition image $\mathbf{y}$
\begin{equation}
\begin{aligned}
p_\theta\left(\mathbf{x}_{0: T-1} \mid \mathbf{x}_T, \mathbf{y}\right) & =p\left(\mathbf{x}_T\right) \prod_{t=1}^T p_\theta\left(\mathbf{x}_{t-1} \mid \mathbf{x}_t, \mathbf{y}\right) \\
p_\theta\left(\mathbf{x}_{t-1} \mid \mathbf{x}_t, \mathbf{y}\right) & =\mathcal{N}\left(\mathbf{x}_{t-1} \mid \mu_\theta\left(\mathbf{y}, \mathbf{x}_t, t\right), \sigma^{2}_t \mathbf{I}\right),
\end{aligned}
\end{equation}
where the mean $\mu_\theta\left(\mathbf{y},\mathbf{x}_t, t\right)$ is learnt with a Denoising UNet with model parameters $\theta$ \citep{ho2020denoising}, and the variance is fixed to $\sigma^{2}_t=\frac{1-\gamma_{t-1}}{1-\gamma_{t}}\beta_{t}$ as suggested by \citet{peng2022towards}. This reverse process can be regarded as a Markov chain with learned Gaussian transitions with an initial state of $p(\mathbf{x}_T)=\mathcal{N}(0, \mathbf{I})$. However, to approximate the reverse process with this network parameterized Gaussian distribution, the value of $\beta_t$ must be sufficiently constrained to ensure that the forward and reverse processes have approximately the same functional form \citep{sohl2015deep,ho2020denoising}. As a result, $T$ should be selected large enough (typically thousands of diffusion steps) to make $p\left(\mathbf{x}_T\right)$ virtually identical to standard Gaussian noise $\mathcal{N}(0, \mathbf{I})$. This requirement makes the reverse diffusion process, or the sampling process, extremely slow due to the thousands of network evaluations. To overcome this issue, taking inspiration from \citet{zheng2022truncated}, we shorten the sampling process by truncating the diffusion chain at $T_{trunc}$ and initiating the reverse denoising from a noisy image $\mathbf{x}_{T_{trunc}}$ rather than pure Gaussian noise, as depicted in Fig.\ref{fig1}(b). The truncated denoising diffusion process is now characterized as

\begin{equation}
p_\theta\left(\mathbf{x}_{0: T_{trunc}-1} \mid \mathbf{x}_{T_{trunc}}, \mathbf{y}\right) =p\left(\mathbf{x}_{T_{trunc}}\right) \prod_{t=1}^{T_{trunc}} p_\theta\left(\mathbf{x}_{t-1} \mid \mathbf{x}_t, \mathbf{y}\right)
\label{eqn4}
\end{equation}
The distribution of noisy images at the truncation point $p\left(\mathbf{x}_{T_{trunc}}\right)$ is estimated from Gaussian noise using a normalizing flow-based network, which will be detailed in Section \ref{sec_flow}. To fulfill Equation \ref{eqn4}, a Denoising UNet $\epsilon_\theta$ with parameters $\theta$ is trained to estimate the noise $\epsilon$ in $\mathbf{x}_t$ at a given timestep $t$
\begin{equation}
\mathbb{E}_{(\mathbf{x_0}, \mathbf{y})} \mathbb{E}_{\epsilon, t, s}\left\|\epsilon-\epsilon_\theta\left(\mathbf{y},\mathbf{x}_t, t, s\right)\right\|_2^2
\end{equation}
where $\mathbf{x}_t=\sqrt{\gamma_t} \mathbf{x}_{0} + (1-\gamma_t)\epsilon$ and $\epsilon \sim \mathcal{N}(0, \mathbf{I})$ based on Equation \ref{eqn2}. Timestep $t$ is uniformly sampled from $\mathcal{U}(1, T_{trunc})$. $s$ represents the upscaling factor, which will be detailed in Section \ref{sec_multiscale}.

The detailed architecture of the Denoising UNet is shown in Fig.\ref{fig2}, adhering to the design presented by \citet{nichol2021improved}. Specifically, $\mathbf{x}_0$ is a single noiseless high-resolution MRSI metabolic map. The condition $\mathbf{y}$ is a concatenation of four components: low-resolution (LR) metabolic map, quality-filtering mask $M$, T1 MRI and FLAIR MRI, i.e. $\mathbf{y}=\{$LR, $M$, T1, FLAIR$\}$. The inclusion of the quality-filtering mask enables the network to differentiate between artificial features (voxels with zero intensity) generated by quality filtering and genuine metabolic characteristics. T1 and FLAIR MRIs may offer high-resolution anatomical information beneficial for the super-resolution process \citep{dong2021high}. 

\subsection{Multi-scale Super-resolution}
\label{sec_multiscale}
Upscaling factor is defined as the factor that resolution is improved on each dimension. For example, improving an image's resolution from 16$\times$16 to 64$\times$64 gives an upscaling factor of 4. Depending on the acquisition protocol and the specific MRSI application, experimentally acquired resolutions can vary. Therefore, the upscaling factor often remains unknown prior to training the super-resolution network \citep{dong2022multi}. Most of the previous MRSI super-resolution networks were designed single-scale \citep{iqbal2019super,dong2021high,dong2022flow}, which requires training an independent network for each upscaling factor. This is time-consuming and memory-inefficient. To enable multi-scale super-resolution, we condition the Denoising UNet with the upscaling factor by injecting $s$ to the network. Specifically, the feature maps after the convolution layers in the residual blocks are modulated with Conditional Instance Normalization \citep{huang2017arbitrary}
\begin{equation}
    f' = \hat{\sigma}(t, s)(\frac{f-\mu(f)}{\sigma(f)}) + \hat{\mu}(t, s)
\label{eqn_CIN}
\end{equation}
where $\mu(f)\in\mathbb{R}^{C}$ and $\sigma(f)\in\mathbb{R}^{C}$ are the channel-wise mean and standard deviation of the feature map $f$ with channel number $C$. The modulated mean $\hat{\mu}$ and standard deviation $\hat{\sigma}$ are learned from $t$ and $s$ using a multilayer perceptron (4 fully-connected layers with Sigmoid Linear Units \citep{elfwing2018sigmoid}, or SiLUs, activation functions in between). This approach ensures the network uniquely addresses each upscaling factor.

\subsection{Flow-based Noisy Image Generation}
\label{sec_flow}
Normalizing flow constructs an unknown target distribution from a simple base distribution using a flow of invertible transformations \citep{rezende2015variational}. Given that $\mathbf{x}_{T_{trunc}}$ represents a sample from the distribution of noisy images at the truncation point and $\mathbf{y}$ is the corresponding set of condition images, we use an invertible neural network $F_{\phi}$ to transform $\mathbf{x}_{T_{trunc}}$ into a latent Gaussian variable $\mathbf{z}$, i.e. $\mathbf{z}=F_{\phi}(\mathbf{x}_{T_{trunc}},\mathbf{y})$. Once this transformation is learned, the network can generate samples from the target distribution $p\left(\mathbf{x}_{T_{trunc}}\right)$ by sampling $\mathbf{z}$ from its distribution $p_{\mathbf{z}}(\mathbf{z})$ and inversely passing through the network $\mathbf{x}_{T_{trunc}}=F^{-1}_{\phi}(\mathbf{z},\mathbf{y})$. Based on the change of variable formula, the target distribution can be expressed as

\begin{equation}
    p_{\mathbf{x}_{T_{trunc}}|\mathbf{y}}(\mathbf{x}_{T_{trunc}}|\mathbf{y}) = p_{\mathbf{z}}(\mathbf{z})\left|\text{det}\left(\frac{\partial F_{\phi}(\mathbf{x}_{T_{trunc}}, \mathbf{y})}{\partial \mathbf{x}_{T_{trunc}}}\right)\right|
\end{equation}
with $\frac{\partial F_{\phi}}{\partial \mathbf{x}}$ being the Jacobian matrix \citep{winkler2019learning}. This expression can be reformulated into a negative log-likelihood (NLL) loss for network training

\begin{equation}
    \mathcal{L}_{NLL} = -\text{log} \; p_{\mathbf{z}}(\mathbf{z}) - \text{log} \; \left|\text{det}\left(\frac{\partial (F_{\phi}(\mathbf{x}_{T_{trunc}}, \mathbf{y})}{\partial \mathbf{x}_{T_{trunc}}}\right)\right|
\label{loss_NLL}
\end{equation}
where the first term equals $-\frac{1}{2}(\left\lVert\frac{\mathbf{z} - \boldsymbol{\mu_z}}{\boldsymbol{\sigma_z}}\right\rVert_{2}^{2} + \left\lVert\text{log}(2\pi \boldsymbol{\sigma_z}^2)\right\rVert_{1})$ because $\mathbf{z}$ follows a Gaussian distribution with learned mean $\boldsymbol{\mu_z}$ and standard deviation $\boldsymbol{\sigma_z}$, i.e. $\mathbf{z} \sim \mathcal{N}(\boldsymbol{\mu_z},\boldsymbol{\sigma_z})$.

Observing that the noisy image $\mathbf{x}_{T_{trunc}}$ should center around $\mathbf{x}_0$, we add a guide loss in the sampling direction. This encourages the network to generate an estimated noisy image $\hat{\mathbf{x}}_{T_{trunc}}$ that centers around $\mathbf{x}_0$ \citep{liang2021hierarchical,dong2022flow}. The guide loss is composed of two fidelity-oriented loss functions, pixel loss and structural loss:

\begin{equation}
    \mathcal{L}_{guide} = (1-\alpha) \mathcal{L}_{pixel}(\mathbf{x}_0, \hat{\mathbf{x}}_{T_{trunc}}^{\tau_{f}=0}) + \alpha \mathcal{L}_{structural}(\mathbf{x}_0, \hat{\mathbf{x}}_{T_{trunc}}^{\tau_{f}=0}).
\label{loss_guide}
\end{equation}
$\mathcal{L}_{pixel}$ is the pixel-wise L1-distance between two images, and $\mathcal{L}_{structural}$ measures the Multiscale Structural Similarity (MSSSIM) between two images \citep{dong2021high,wang2003multiscale}. The scalar $\alpha \in (0, 1)$ defines the weight between the two losses. $\tau_{f} \in (0, 1)$ acts as a temperature parameter that controls the variance of the sample: $\hat{\mathbf{x}}_{T_{trunc}}^{\tau_{f}} = F^{-1}_{\phi}(\mathbf{z}, \mathbf{y})$ with $\mathbf{z} \sim \mathcal{N}(\boldsymbol{\mu_z}, \tau_{f}\boldsymbol{\sigma_z})$. In the guide loss, we set $\tau_{f}=0$ to encourage that the center of the distribution of $\hat{\mathbf{x}}_{T_{trunc}}$ locates around $\mathbf{x}_0$.

The overall loss for the flow-based noisy image generation network is a weighted combination of the NLL loss (Equation \ref{loss_NLL}) and the guide loss (Equation \ref{loss_guide})

\begin{equation}
    \mathcal{L}_{flow} = \mathcal{L}_{NLL} + \lambda\mathcal{L}_{guide}
\end{equation}

\subsection{Sampling from FTDDM}
\label{sec_sampling}

To generate noisy images at the truncation point, we use the trained flow-based noisy image generation network $F_{\phi}$:
\begin{equation}
\hat{\mathbf{x}}_{T_{trunc}} = F^{-1}_{\phi}(\mathbf{z}, \mathbf{y}), \quad \mathbf{z} \sim \mathcal{N}(\boldsymbol{\mu_z}, \tau_{f}\boldsymbol{\sigma_z}).
\end{equation}

To generate clean high-resolution MRSI images $\mathbf{x}_0$ from the estimated $\mathbf{x}_{T_{trunc}}$ through the reverse diffusion process, we use the trained Denoising UNet $\epsilon_\theta$ and apply the sampling formula derived in the original DDPM paper \citep{ho2020denoising} iteratively:

\begin{equation}
\mathbf{x}_{t-1}=\frac{1}{\sqrt{1-\beta_t}}\left(\mathbf{x}_t-\frac{\beta_t}{\sqrt{1-\gamma_t}} \epsilon_\theta\left(\mathbf{y}, \mathbf{x}_t, t, s\right)\right)+\sigma_t \mathbf{z}_t
\end{equation}
where $t=T_{trunc}, \ldots, 1$, and $\mathbf{z}_t$ is a latent Gaussian variable for the diffusion model at timestep $t$. Observing that the visual sharpness of the generated images is correlated to the variance of the latent Gaussian variable in the flow-based networks \citep{dong2022flow, lugmayr2020srflow}, we analogously introduce a temperature parameter $\tau_{d} \in (0, 1)$ into our diffusion model, such that $\mathbf{z}_t \sim \mathcal{N}(0, \tau_{d}\mathbf{I})$. In this way, the sharpness of the generated images can be flexibly adjusted by $\tau_{d}$. The overall sampling process of FTDDM is described in Algorithm \ref{algo}.

\begin{algorithm}
\caption{Sampling from FTDDM}\label{alg}
\begin{algorithmic}[1]
\State $\mathbf{z} \sim \mathcal{N}(\boldsymbol{\mu_z}, \tau_{f}\boldsymbol{\sigma_z})$
\State $\hat{\mathbf{x}}_{T_{trunc}} = F^{-1}_{\phi}(\mathbf{z}, \mathbf{y})$
\State $\mathbf{x}_{T_{trunc}} = \hat{\mathbf{x}}_{T_{trunc}}$
\For{$t=T_{trunc}, \ldots, 1$}
    \State $\mathbf{z}_t \sim \mathcal{N}(0, \tau_{d}\mathbf{I}) \text { if } t>1 \text {, else } \mathbf{z}_t=0$
    \State $\mathbf{x}_{t-1}=\frac{1}{\sqrt{1-\beta_t}}\left(\mathbf{x}_t-\frac{\beta_t}{\sqrt{1-\gamma_t}} \epsilon_\theta\left(\mathbf{y},\mathbf{x}_t,t,s\right)\right)+\sigma_t \mathbf{z}_t$
\EndFor
\State return $\mathbf{x}_0$
\end{algorithmic}
\label{algo}
\end{algorithm}

\section{Experiments and Results}

\subsection{Implementation Details}
We implemented 5-fold cross-validation on our dataset comprising 25 patients. In each fold, 15 patients were used for training, 5 for validation and another 5 for testing. Across these folds, the average counts of 2D metabolic maps used for training, validation and testing are 1365, 455 and 455, respectively. The training dataset was substantially augmented using random rotation, flipping and translation during training. The validation dataset was used to monitor convergence and save the model with the lowest validation loss. The low resolutions were uniformly sampled during training from 13 values: LR $\in\{$8×8, 10×10, 12×12, ..., 30×30, 32×32$\}$, which corresponds to upscaling factors of $s\in\{$8.0, 6.4, 5.3, ..., 2.1, 2.0$\}$. T1 and FLAIR images were down-sampled to a consistent resolution of 64$\times$64 with the high-resolution metabolic maps and were normalized between [0, 1] before being input to the network.

The Denoising UNet, shown in Fig.\ref{fig2} follows the same architecture as proposed in \citet{nichol2021improved}. It has four stages of downsampling and four stages of upsampling, each with three convolutional residual blocks. The number of channels used at each stage is 64, 128, 192 and 256, respectively. The upscaling factor $s$ and sinusoidal embedding of timestep $t$ are passed to the residual blocks via Conditional Instance Normalization. Multi-head attention blocks with four attention heads are used between the residual blocks at 16$\times$16 and 8$\times$8 feature map resolution levels.

The diffusion models employed a total of $T$=1000 diffusion steps, as recommended by \citet{ho2020denoising}. Our FTDDM truncated the diffusion chain at $T_{trunc}$=100, equivalent to roughly 10-fold acceleration. $\beta_t$ follows a cosine noise schedule as proposed by \citet{nichol2021improved}. The temperature parameters for sampling from the diffusion model and the flow-based network were set to $\tau_{d}$=0.9 and $\tau_{f}$=0.9, respectively. The weighting parameters for the loss function are $\alpha$=0.84 \citep{zhao2016loss} and $\lambda$=10 \citep{dong2022flow}. The proposed model was trained with Adam optimizer \citep{kingma2014adam}, 100,000 iterations and batch size of 8. The initial learning rate was set to 1×10$^{-4}$, followed by a decay per iteration. The methods for comparison were implemented based on their original publications and also trained in a multi-scale manner. All models were implemented with PyTorch and run on NVIDIA GTX 1080 and V100 GPUs.

\subsection{Experimental Studies}

\begin{table*}[!t]
\setlength{\tabcolsep}{4.4pt} 
\renewcommand{\arraystretch}{1.2} 
\scriptsize
\caption{Quantitative comparisons of FTDDM against other deep learning methods at three upscaling factors $s$=8.0, 4.0, 2.0. NFE: number of function evaluations, Time: Sampling time in seconds (s) for a single image slice, Params: number of model parameters in millions (M), NRMSE: Normalized
Root Mean Square Error, PSNR: Peak Signal-to-Noise Ratio, SSIM: Structural Similarity Index.}
\centering
\begin{tabular}{p{2.95cm} | M{0.45cm} | M{0.5cm} | M{0.7cm} | M{1.2cm}M{0.8cm}M{0.9cm} c | M{1.2cm}M{0.8cm}M{0.9cm} c | M{1.2cm}M{0.8cm}M{1.1cm}}
\hline
 & & & & \multicolumn{3}{c}{$s$=8.0 (LR=8×8)} && \multicolumn{3}{c}{$s$=4.0 (LR=16×16)} && \multicolumn{3}{c}{$s$=2.0 (LR=32×32)}\\
\cline{5-8} \cline{9-12} \cline{13-15}
 Methods & NFE & Time (s) & Params (M) & $\;\;$NRMSE$\downarrow$ & $\;$PSNR$\uparrow$ & $\quad$SSIM$\uparrow$ && $\;\;$NRMSE$\downarrow$ & $\;$PSNR$\uparrow$ & $\quad$SSIM$\uparrow$ && $\;\;$NRMSE$\downarrow$ & $\;$PSNR$\uparrow$ & $\quad$SSIM$\uparrow$ \\
\hline
MUNet-FS-cWGAN & 1 & 0.05 & 27.5 & 0.414$\pm$0.188 & 26.0$\pm$2.3 & 0.864$\pm$0.043 && 0.344$\pm$0.164 & 27.7$\pm$2.5 & 0.904$\pm$0.036 && 0.238$\pm$0.110 & 30.9$\pm$2.8 & 0.956$\pm$0.019 \\
Flow Enhancer & 1 & 0.22 & 37.6 & 0.380$\pm$0.193 & 26.9$\pm$2.4 & 0.882$\pm$0.038 && 0.321$\pm$0.175 & 28.5$\pm$2.5 & 0.917$\pm$0.031 && 0.240$\pm$0.140 & 31.1$\pm$2.7 & 0.956$\pm$0.020 \\
DDPM ($T$=1000) & 1000 & 12.44 & 30.3 & 0.364$\pm$0.174 & 27.2$\pm$2.5 & 0.884$\pm$0.036 && 0.287$\pm$0.136 & 29.3$\pm$2.9 & 0.927$\pm$0.027 && 0.183$\pm$0.086 & 33.2$\pm$3.4 & 0.970$\pm$0.014 \\
DDPM ($T_{respace}$=100) & 100 & 1.25 & 30.3 & 0.395$\pm$0.183 & 26.4$\pm$2.3 & 0.868$\pm$0.040 && 0.317$\pm$0.145 & 28.3$\pm$2.6 & 0.911$\pm$0.031 && 0.212$\pm$0.096 & 31.9$\pm$2.9 & 0.958$\pm$0.018 \\
DPM-Solver++ ($T_{solver}$=100) & 101 & 1.46 & 30.3 & 0.413$\pm$0.183 & 26.0$\pm$2.5 & 0.864$\pm$0.041 && 0.325$\pm$0.147 & 28.1$\pm$2.9 & 0.911$\pm$0.032 && 0.208$\pm$0.094 & 32.1$\pm$3.4 & 0.962$\pm$0.017 \\
TDPM ($T_{trunc}$=100) & 101 & 1.27 & 60.6 & 0.382$\pm$0.169 & 26.6$\pm$2.6 & 0.876$\pm$0.038 && 0.290$\pm$0.138 & 29.2$\pm$3.0 & 0.927$\pm$0.028 && 0.175$\pm$0.082 & 33.6$\pm$3.7 & 0.972$\pm$0.014 \\
\textbf{FTDDM} ($T_{trunc}$=100) & 101 & 1.33 & 50.5 & \textbf{0.356$\pm$0.168} & \textbf{27.3$\pm$2.5} & \textbf{0.888$\pm$0.036} && \textbf{0.276$\pm$0.134} & \textbf{29.6$\pm$2.9} & \textbf{0.932$\pm$0.027} && \textbf{0.170$\pm$0.080} & \textbf{33.9$\pm$3.6} & \textbf{0.974$\pm$0.013} \\
\hline
\end{tabular}
\label{table1}
\end{table*}

We conducted both quantitative and qualitative comparisons of our proposed method, FTDDM, against other state-of-the-art deep learning models. MUNet-FS-cWGAN \citep{dong2022multi} is a Multi-encoder UNet (MUNet) architecture trained with conditional Wasserstein GAN (cWGAN). It uses a Filter-Scaling (FS) strategy for multi-scale super-resolution. The Flow Enhancer \citep{dong2022flow} is a normalizing flow-based invertible network designed to enhance the visual quality of the blurry super-resolution images given by MUNet. DDPM \citep{ho2020denoising} is the conventional diffusion model that uses the full diffusion chain $T$=1000 to iteratively transform Gaussian noise into high-resolution MRSI images (Fig.\ref{fig1}(a)), given the condition images $\mathbf{y}$. As a baseline method to reduce sampling time, we implemented the simple striding, or respacing, method to reduce DDPM sampling steps from $T=1000$ to $T_{respace}$=100 by evenly skipping the diffusion steps \citep{nichol2021improved}. DPM-Solver++ is a high-order ODE solver designed to generate high-quality samples from diffusion models within a small number of steps \citep{lu2022dpm}, which we have implemented for DDPM. Furthermore, we implemented TDPM \citep{zheng2022truncated} which uses a truncated diffusion chain and adopts a GAN to estimate the noisy images at the truncation point. For quantitative analyses, we evaluated the super-resolution performance using three metrics, Normalized Root Mean Square Error (NRMSE), Peak Signal-to-Noise Ratio (PSNR) and Structural Similarity Index (SSIM). Sampling efficiency was measured by the number of function evaluations (NFE) \citep{zheng2022truncated}.

Our FTDDM consists of two components: a diffusion model (Truncated Denoising Diffusion) and a normalizing flow model (Flow-based Noisy Image Generation). We conducted ablation studies to showcase the significance of each component. Key design elements, including the guide loss $\mathcal{L}_{guide}$ and the uses of quality-filtering mask $M$ and MRIs as condition images, were also justified through ablation studies. We studied the selection for $T_{trunc}$, testing the performance of FTDDM at values of $T_{trunc}$=20, 50 and 100. We also attempted to combine DPM-Solver++ with FTDDM to further reduce the number of sampling steps, observing that the remaining sampling steps in the Truncated Denoising Diffusion model can be resolved using DPM-Solver++. Moreover, to confirm that conditioning on the upscaling factor using Equation \ref{eqn_CIN} aids the network in processing each low resolution at the right scale, we studied the network performance across different combinations of actual and conditioned upscaling factors. Furthermore, effect of the temperature parameter $\tau_d$ on the image sharpness was investigated. The image sharpness, or visual quality, was evaluated quantitatively using Learned Perceptual Image Patch Similarity (LPIPS), which measures the high-level similarity between two images using a pretrained VGG network and aligns closely with human perceptual judgment \citep{zhang2018unreasonable}. 

To confirm the clinical advantage of our method, we invited 2 experienced neuroradiologists to assess the quality of our super-resolution images relative to ground truth from clinical perspectives. Radiologists' rating was considered as a key evaluation criterion in previous deep learning-based image reconstruction tasks \citep{muckley2021results,dong2022invertible,zhou2022dual}. To determine the lowest resolution the network can effectively manage, or the ``breakpoint" of our method, we asked the neuroradiologists to evaluate the model performance at three extremely low input resolutions of 6$\times$6, 4$\times$4 and 2$\times$2.

\subsection{Results and Discussion}

\begin{figure*}[t!]
\centering
\includegraphics[width=0.9\textwidth]{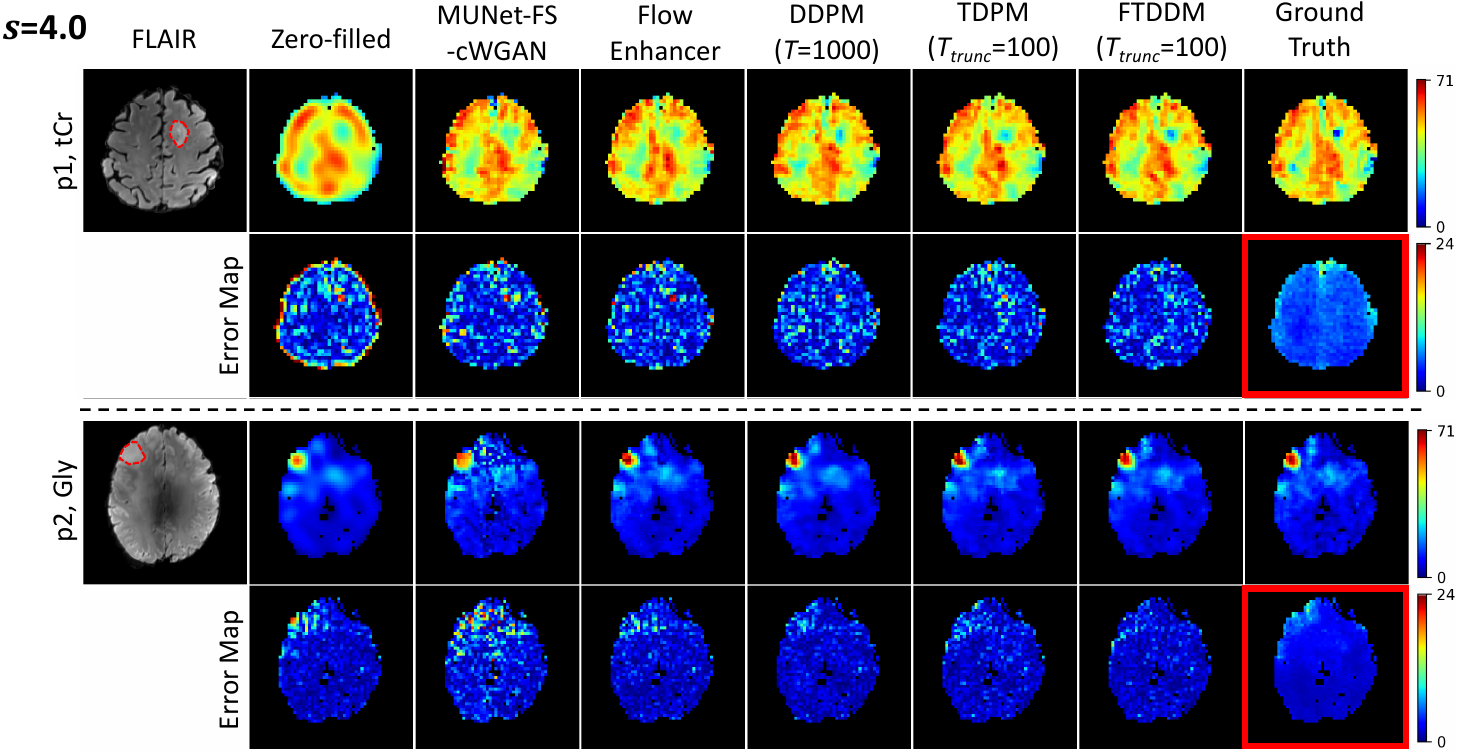}
\caption{Qualitative comparisons of FTDDM against other methods at upscaling factor $s$=4.0. The two examples are: a tCr image from patient p1 and a Gly image from patient p2. FLAIR MRI provides the corresponding anatomical reference, with the tumor delineated by the red dashed line. Each metabolic map is shown alongside with its error map, except for ground truth. Note that the images below ground truth, framed in red, are the standard deviation maps of 50 FTDDM samples and can be used for uncertainty estimation (they are not error maps of ground truth).}
\label{fig_main}
\end{figure*}

\begin{figure*}[t!]
\centering
\includegraphics[width=0.9\textwidth]{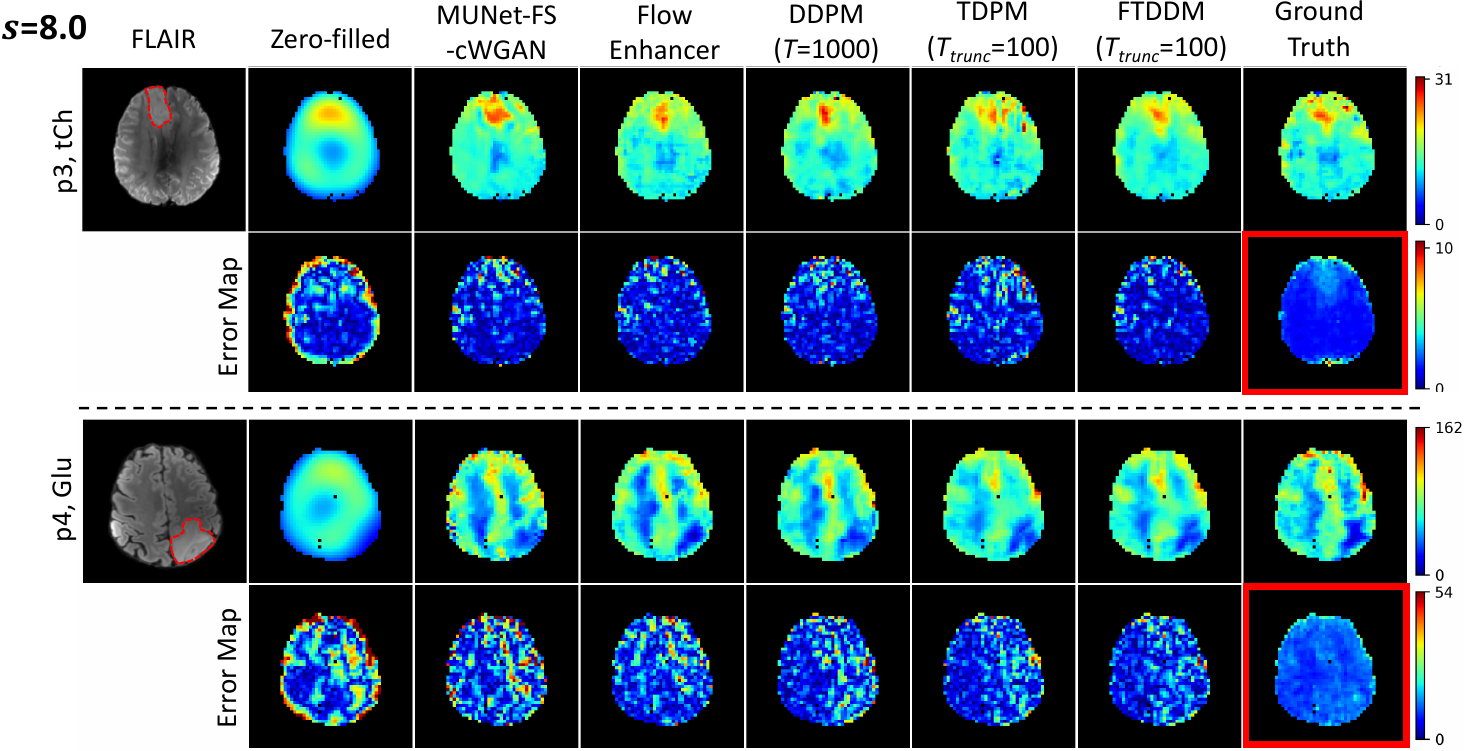}
\caption{Qualitative comparisons of FTDDM against other methods at upscaling factor $s$=8.0. The two examples are: a tCh image from patient p3 and a Glu image from patient p4.}
\label{fig_main2}
\end{figure*}

\subsubsection{Quantitative evaluations}
\label{sec_quantitative}
Table \ref{table1} presents the quantitative comparison of the proposed FTDDM against other deep learning models at three different upscaling factors $s$=8.0, 4.0 and 2.0. Compared to the GAN-based model (MUNet-FS-cWGAN) and the flow-based model (Flow Enhancer), the diffusion model DDPM excels across all metrics. This highlights the superior learning capability of the diffusion model within this set of generative models. However, DDPM requires 1000 NFE during sampling, translating to a sampling time of approximately 12.44 seconds for a single image slice on our devices. Reducing the NFE to 100 with the simple respacing method ($T_{respace}$=100) downgrades the performance, as skipping the diffusion steps yields images that are less denoised. Using the same number of sampling steps, DPM-Solver++ with $T_{solver}$=100 fails to perform better than the respacing method. However, we show in Section \ref{sec_selectT} that DPM-Solver++ can outperform the respacing method when fewer sampling steps are used. The truncation method TDPM at $T_{trunc}$=100 outperforms the respacing method and offers comparable performance with the DDPM at $T$=1000. TDPM requires 101 NFE, due to an extra GAN function evaluation. Our proposed method, FTDDM, achieves even better performance than TDPM, indicating the flow-based network's superiority over GAN in generating noisy images at the truncation point. For a single image slice, FTDDM at $T_{trunc}$=100 has a sampling time of approximately 1.33 seconds, which is over 9-fold acceleration relative to DDPM. Since evaluating the flow-based network for generating noisy images at the truncation point takes extra time (one additional NFE), the speed increase is slightly below 10-fold. Wilcoxon signed-rank tests were conducted between each pair of scores for all methods in Table \ref{table1}, for which the results indicate that all differences are significant with P values $<$ 0.001.

\subsubsection{Qualitative evaluations}
\label{sec_qualitative}
The quantitative improvement given by FTDDM is reflected by the better image quality and tumor visualization in qualitative analysis. The qualitative results of different methods at two different upscaling factors $s$=4.0 and $s$=8.0 are displayed in Fig.\ref{fig_main} and Fig.\ref{fig_main2}, respectively. The low-resolution images are interpolated to the same resolution as the high-resolution images (64$\times$64) by filling the high-frequency components of the k-space with zeros, i.e. zero-filled images. Compared to the ground truth high-resolution images, the zero-filled low-resolution images appear blurry and lack high-frequency image details. While MUNet-FS-cWGAN and Flow Enhancer can recover a large portion of high-frequency features, they either fail to capture the tumor contrast (tCr map from Fig.\ref{fig_main}) or distort the tumor shape (Gly map from Fig.\ref{fig_main} and both maps from Fig.\ref{fig_main2}) when compared to the ground truth. DDPM has greater learning capability and reconstructs the tumors more consistent with the ground truth. Additionally, images given by DDPM more accurately capture the characteristics of white matter and gray matter. TDPM, due to its GAN dependence, can exhibit training instability and occasionally introduce pronounced artifacts, as observed in the tCh map from Fig.\ref{fig_main2}. Among the diffusion models, FTDDM reconstructs tumor characteristics closest to ground truth and yields the lowest overall error. The error maps of FTDDM exhibit less bright spots compared to the other methods. The uncertainty map (framed in red) for FTDDM was calculated from the standard deviation of 50 repeated samples and acts as a tool for error estimation. Typically, uncertainty is more pronounced around the brain's periphery, suggesting the model detected a higher variance in these areas. This observation aligns with \citet{de2019vivo} and \citet{ziegs2023test}, which mentions the stronger signal distortion due to extracranial lipid contamination and B$_0$ field inhomogeneity proximal to the skull.

\subsubsection{Ablation studies}
Table \ref{tab2} shows the ablation studies on the diffusion model and the normalizing flow model in FTDDM at an upscaling factor of $s$=4.0. FTDDM is equivalent to a normalizing flow model when the truncation point is selected as $T_{trunc}$=0 and is equivalent to a diffusion model when $T_{trunc}$=1000. Using the flow model alone underperforms the combined model by a large margin due to the missing diffusion process. Conversely, using the diffusion model alone requires a large number of sampling steps while still underperforming the combined model. Ablation studies on specific design elements at an upscaling factor of $s$=8.0 are presented in Table \ref{tab3}. Omitting the quality-filtering mask from the condition images makes the model interpret artificial features introduced by quality filtering as genuine image features, resulting in a significant deterioration in model performance. Surprisingly, the removal of T1 and FLAIR MRIs only marginally affects the performance, which suggests that the correlation between MRIs and MRSI is somewhat trivial for our in vivo dataset. The network does not seem to integrate substantial anatomical information from MRIs into the super-resolution MRSI. This implies that the traditional super-resolution MRSI methods that rely heavily on MRI-based regularization \citep{jain2017patch,lam2014subspace,kasten2016magnetic} might fail on in vivo MRSI dataset. Lastly, the guidance loss helps to boost the model performance. Wilcoxon signed-rank tests indicate that all differences in Table \ref{tab2} and Table \ref{tab3} are statistically significant with P values $<$ 0.001.

\begin{table}
\setlength{\tabcolsep}{4.2pt}
\renewcommand{\arraystretch}{1.3} 
\footnotesize
\caption{Ablation studies on the diffusion model and the normalizing flow model in FTDDM.}
\centering
{\begin{tabular}{c|ccc}
\hline
  Method& NRMSE$\downarrow$ & PSNR$\uparrow$ & SSIM$\uparrow$ \\
\hline
Flow only ($T_{trunc}$=0) & 0.320$\pm$0.155 & 28.3$\pm$2.3 & 0.910$\pm$0.030 \\
Diffusion only ($T_{trunc}$=1000)& 0.282$\pm$0.134 & 29.4$\pm$2.9 & 0.929$\pm$0.027 \\
Diffusion + Flow ($T_{trunc}$=100) & \textbf{0.276$\pm$0.134} & \textbf{29.6$\pm$2.9} & \textbf{0.932$\pm$0.027} \\
\hline
\end{tabular}}
\label{tab2}
\end{table}

\begin{table}
\renewcommand{\arraystretch}{1.3} 
\footnotesize
\caption{Ablation studies on design elements. $\checkmark$ or $\times$ indicates whether a certain design element is
included or not.}
\centering
\begin{tabular}{ccc|cccc}
\hline
  Mask & MRI & $L_{guide}$ & NRMSE$\downarrow$ & PSNR$\uparrow$ & SSIM$\uparrow$ \\
\hline
$\times$ & $\checkmark$ & $\checkmark$ & 0.419$\pm$0.221 & 25.9$\pm$2.7 & 0.866$\pm$0.047 \\
$\checkmark$ & $\times$ & $\checkmark$ & 0.361$\pm$0.170 & 27.2$\pm$2.5 & 0.885$\pm$0.036 \\
$\checkmark$ & $\checkmark$ & $\times$ & 0.358$\pm$0.169 & 27.3$\pm$2.4 & 0.884$\pm$0.035 \\
$\checkmark$ & $\checkmark$ & $\checkmark$ & \textbf{0.356$\pm$0.168} & \textbf{27.3$\pm$2.5} & \textbf{0.888$\pm$0.036}  \\
\hline
\end{tabular}
\label{tab3}
\end{table}

\begin{figure}[b]
\centering
\includegraphics[width=8.8cm]{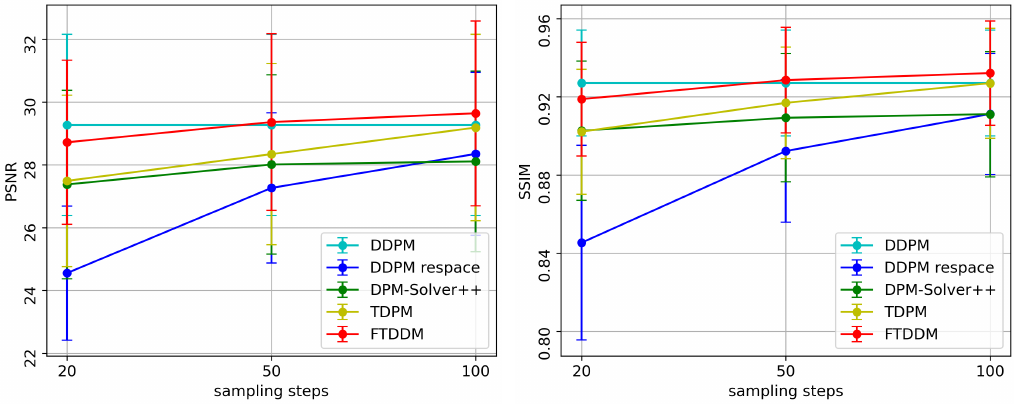}
\caption{Model performance of DDPM, DDPM with respacing, DPM-Solver++, TDPM and FTDDM at different numbers of sampling steps ($T_{respace}$ for DDPM respace, $T_{solver}$ for DPM-Solver++, $T_{trunc}$ for TDPM and FTDDM).}
\label{fig5}
\end{figure}

\begin{table}
\setlength{\tabcolsep}{3.0pt}
\renewcommand{\arraystretch}{1.3} 
\footnotesize
\caption{Study of combining DPM-Solver++ with FTDDM to further reduce the number of sampling steps. SSIM scores are shown for each valid combination of $T_{trunc}$ in FTDDM and $T_{solver}$ in DPM-Solver++.}
\centering
{\begin{tabular}{c|c|ccc}
\hline
  & no DPM-solver++ & $T_{solver}$=50 & $T_{solver}$=20 & $T_{solver}$=10\\
\hline
$T_{trunc}$=100 & 0.932$\pm$0.027 & 0.928$\pm$0.028 & 0.928$\pm$0.028 & 0.927$\pm$0.028\\
$T_{trunc}$=50 & 0.929$\pm$0.027 & - & 0.922$\pm$0.029 & 0.922$\pm$0.029 \\
$T_{trunc}$=20 & 0.919$\pm$0.029 & - & - & 0.912$\pm$0.031 \\
\hline
\end{tabular}}
\label{tab4}
\end{table}

\subsubsection{Selection of $T_{trunc}$}
\label{sec_selectT}
We studied the performance of different models across various sampling steps, specifically, 100, 50 and 20 (Fig.\ref{fig5}). Compared to the unaccelerated DDPM (the cyan line), the respacing method (the blue line) significantly degrades the performance as fewer sampling steps are used ($T_{respace}$=50 and $T_{respace}$=20). Using the ODE solver DPM-Solver++, the rate of this degradation is considerably slower. The truncation method TDPM outperforms DPM-Solver++, and our FTDDM achieves the best performance among these acceleration methods.
At around $T_{trunc}$=50, FTDDM achieves a similar level of performance as the unaccelerated DDPM, which is equivalent to an approximately of 20-fold acceleration. Therefore, although we used $T_{trunc}$=100 throughout this paper to pursue a better performance, our model can certainly be operated at $T_{trunc}$=50 to attain greater acceleration. At $T_{trunc}>$50, our FTDDM is able to outperform DDPM because the Denoising UNet can focus on denoising a less noisy image $\mathbf{x}_{T_{trunc}}$, which is easier to learn than denoising a pure Gaussian noise image $\mathbf{z}$ as in DDPM. This is consistent with the observation that less training iteration is required for the diffusion part to converge when $T_{trunc}$ is smaller \citep{zheng2022truncated}.

\begin{figure}[b!]
\centering
\includegraphics[width=8.8cm]{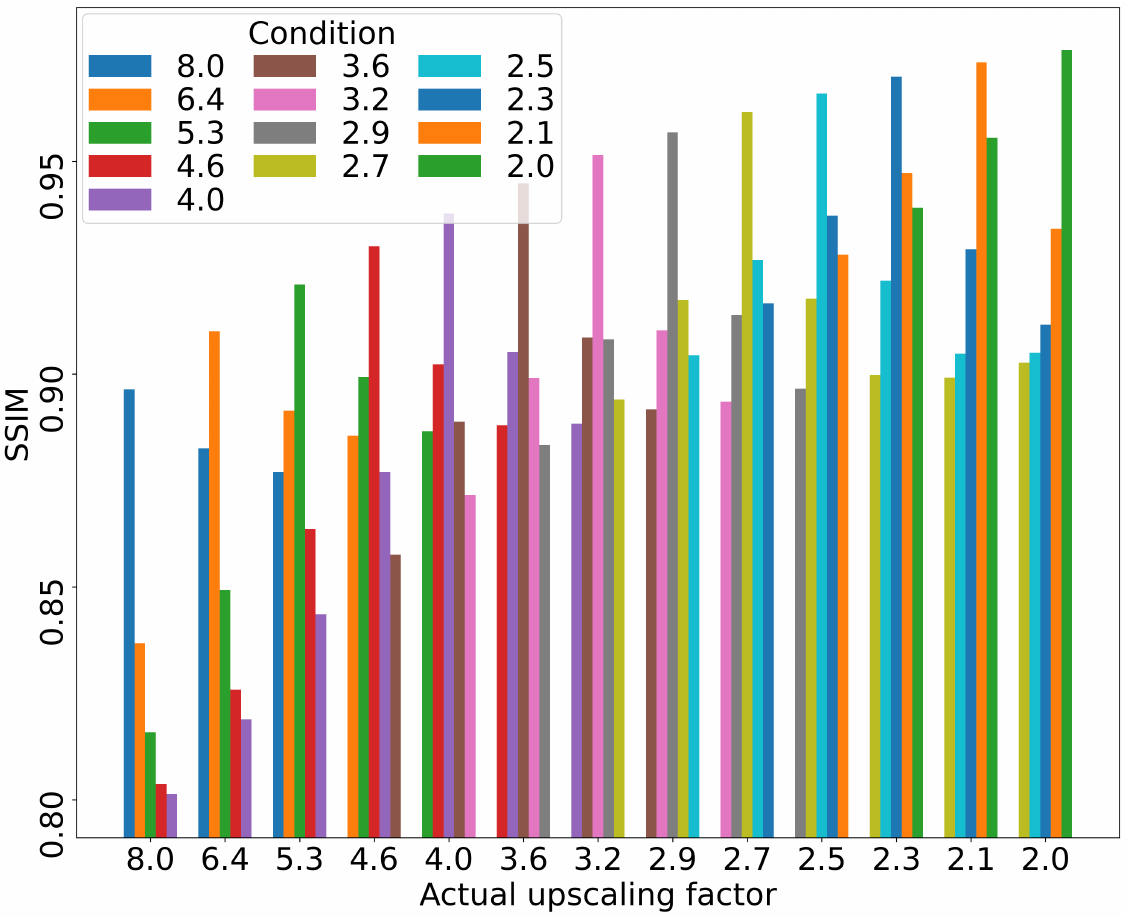}
\caption{Model performance (measured in SSIM) under different combinations of actual upscaling factors (horizontal axis) and conditioned upscaling factors (color bars). For conciseness, only 5 adjacent values of conditions are shown for each actual upscaling factor.}
\label{fig6}
\end{figure}

\begin{figure}[t]
\centering
\includegraphics[width=8.8cm]{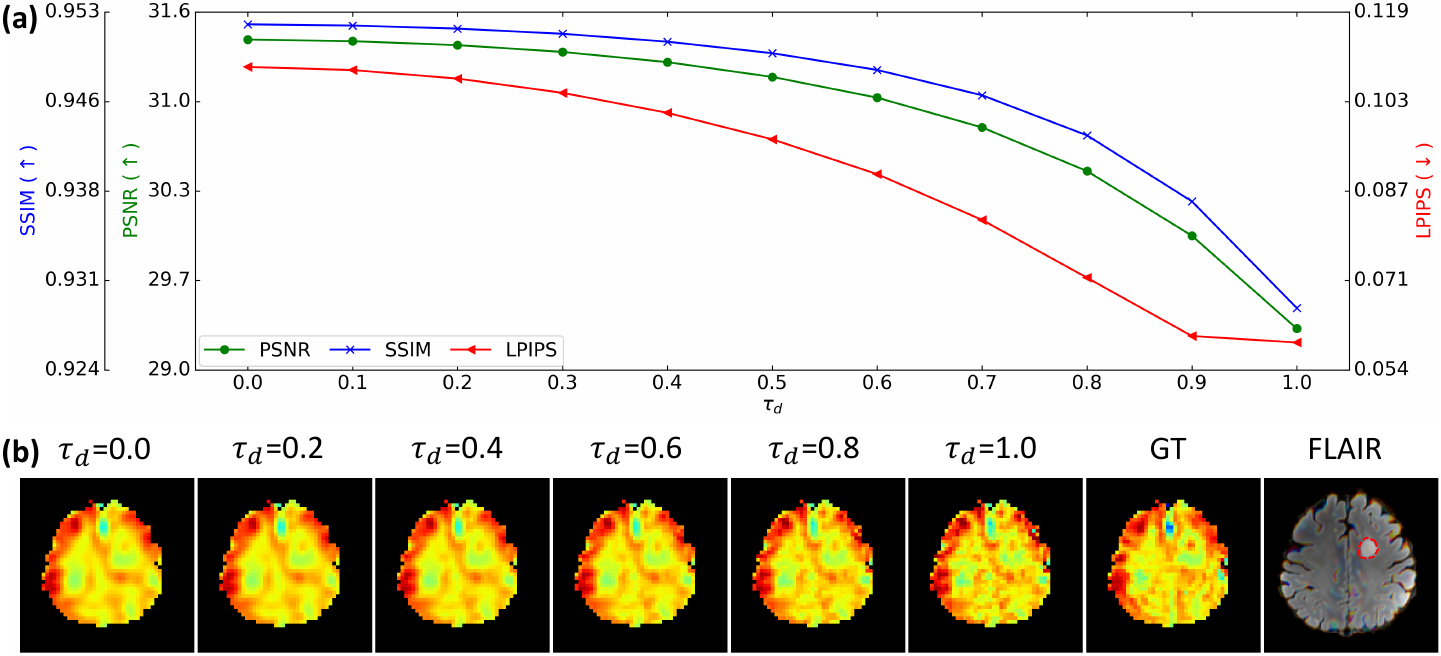}
\caption{Sharpness adjustment with $\tau_d$. (a) Network performance in PSNR, SSIM and LPIPS at various levels of $\tau_d$. Lower LPIPS ($\downarrow$) is better. (b) Visual sharpness of a super-resolution Ins image at various $\tau_d$ compared to ground truth (GT). The tumor is delineated by the red dashed line in FLAIR.}
\label{fig7}
\end{figure}

\begin{figure*}
\centering
\includegraphics[width=0.8\textwidth]{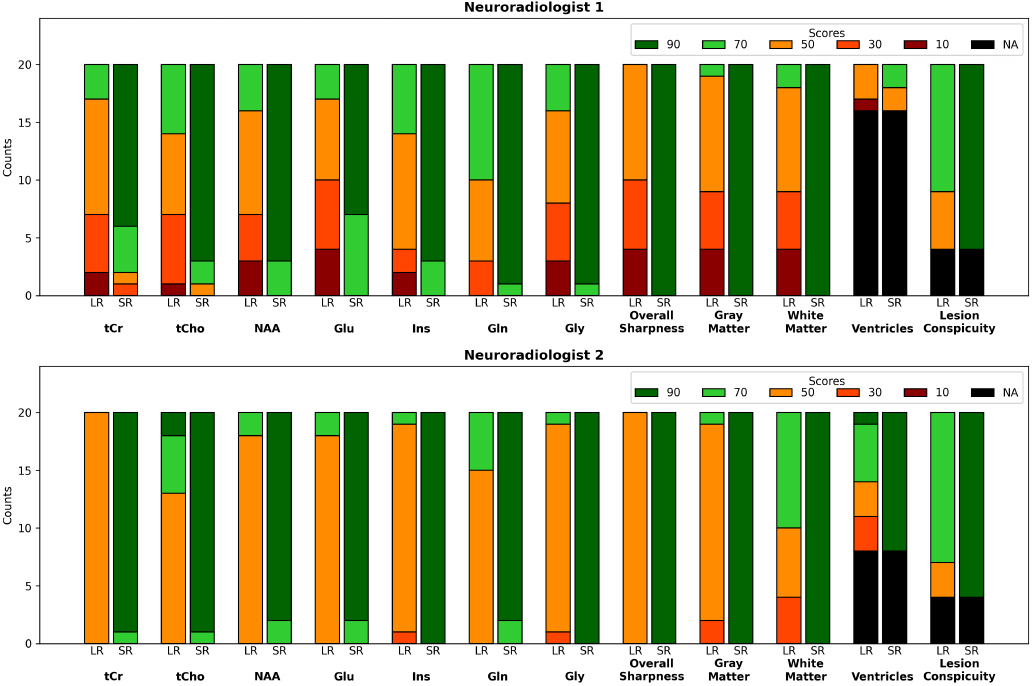}
\caption{Image quality assessment based on neuroradiologists' ratings. Image quality of low-resolution (LR) images and super-resolution (SR) images given by FTDDM was blind-reviewed by Neuroradiologist 1 and 2. The horizontal axis shows the evaluation metrics. Each colored bar denotes the count of images that received a specific score for the corresponding metric. The scoring system is: NA (not applicable), 10 (poor), 30 (bad), 50 (fair), 70 (good) and 90 (excellent).}
\label{fig_rad_exp1}
\end{figure*}

\begin{figure}[t]
\centering
\includegraphics[width=8.8cm]{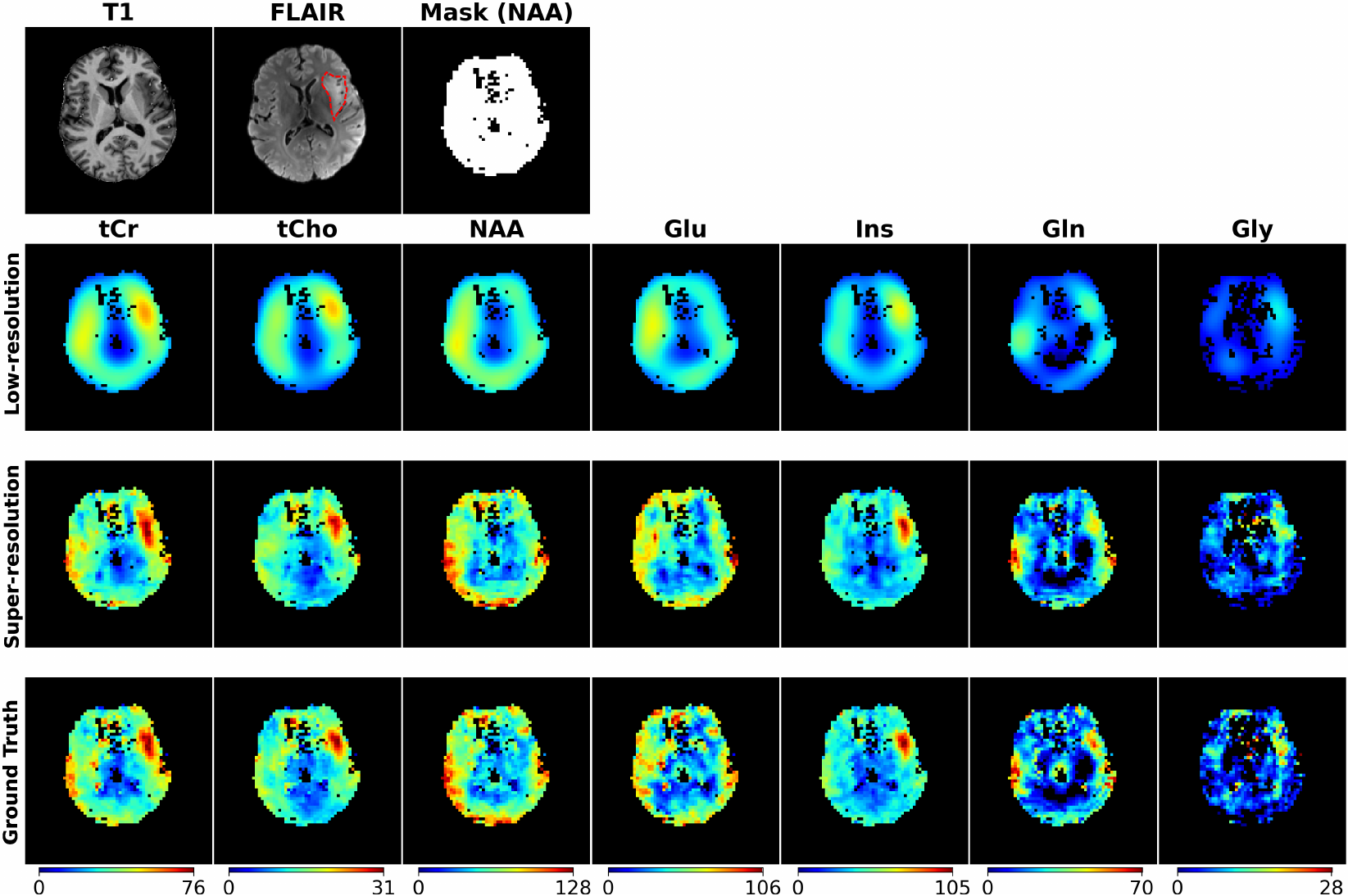}
\caption{Visualization of low-resolution (zero-filled from 8$\times$8), FTDDM super-resolution and ground truth images of 7 metabolites from a slice containing tumor. The first row shows T1 MRI, FLAIR MRI and an example of quality-filtering mask for NAA. The tumor is delineated by the red dashed line in FLAIR.}
\label{fig_allmets}
\end{figure}

\subsubsection{Combining with DPM-Solver++}
Observing that the remaining sampling steps in the Truncated Denoising Diffusion model can be further reduced with DPM-Solver++, we studied how combining DPM-Solver++ with FTDDM impacts the model performance, as shown in Table \ref{tab4}. When $T_{trunc}$=100, the number of sampling steps (or NFE) can be further reduced to 50, 20 or 10 by using DPM-Solver++ at $T_{solver}$=50, 20 or 10. When $T_{trunc}$=50, the number of sampling steps can be further reduced to 20 or 10 by using $T_{solver}$=20 or 10. For $T_{trunc}$=20, it can be reduced to 10. It can be observed that introducing DPM-Solver++ always slightly worsens the performance compared to using FTDDM alone. However, achieving further reduction in sampling steps with fewer $T_{solver}$ only marginally hurts the performance, better than using DPM-Solver++ alone as shown in Fig.\ref{fig5}. Before applying DPM-Solver++ for further reduction, a larger $T_{trunc}$, such as 100, is recommended to ensure a higher upper limit of the performance. 

\begin{figure*}[t]
\centering
\includegraphics[width=0.95\textwidth]{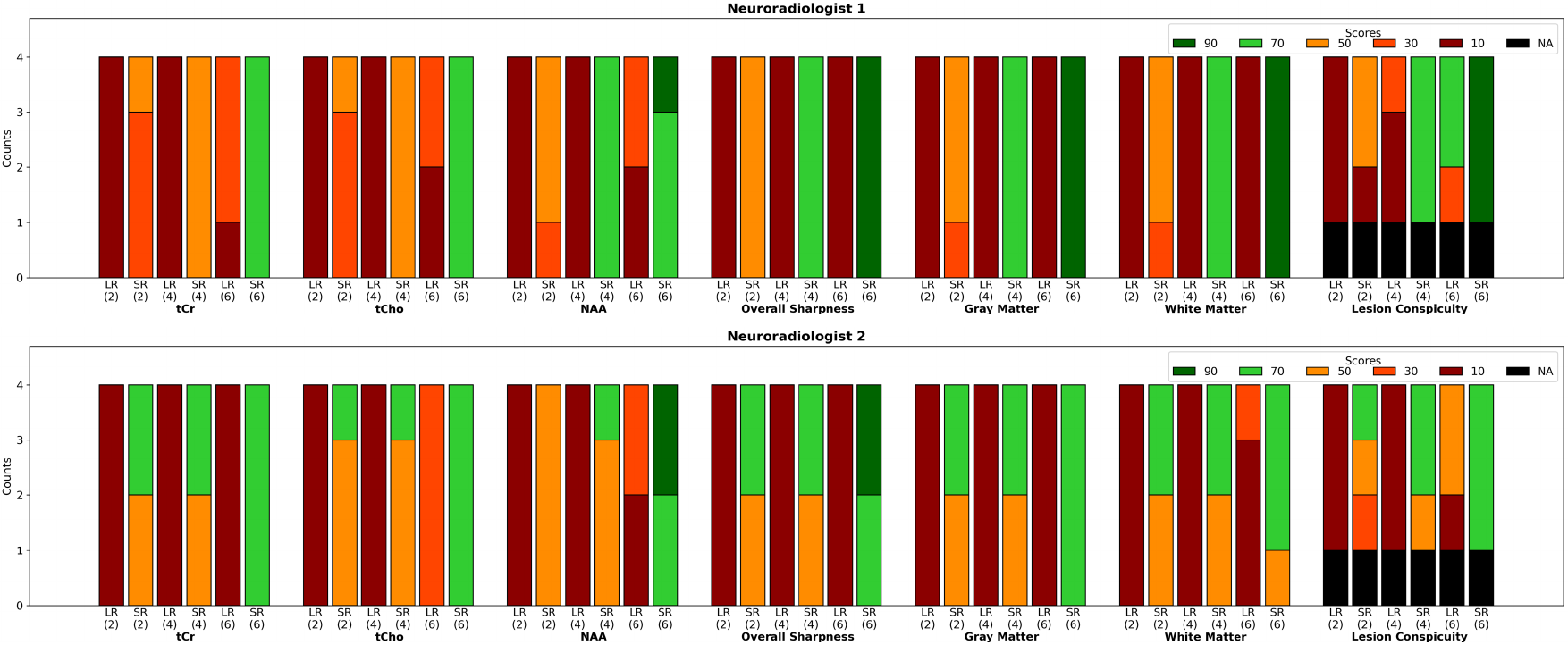}
\caption{Image quality assessment for three extremely low resolutions 2$\times$2, 4$\times$4 and 6$\times$6. Due to limited space, only a subset of most representative evaluation metrics is displayed. For each metric shown on the horizontal axis, ratings for low-resolution (LR) images at 2$\times$2, 4$\times$4 and 6$\times$6 (indicated by the number in the parenthesis) and their corresponding super-resolution (SR) images are presented.}
\label{fig_rad_exp2}
\end{figure*}

\subsubsection{Justification of network conditioning}
We subsequently studied the model performance under various combinations of actual upscaling factors and conditioned upscaling factors, as shown in Fig.\ref{fig6}. The model performance, measured with SSIM, consistently reaches its optimum when the conditioned upscaling factor matches with the actual upscaling factor. This indicates that leveraging Conditional Instance Normalization to modulate the Denoising UNet empowers the network to super-resolve different low resolutions at their appropriate scales.

\subsubsection{Sharpness adjustability}
Our method provides the capability to adjust image sharpness within the same network. The visual sharpness of the super-resolution images can be flexibly controlled by the temperature parameter $\tau_{d}$, which comes with a trade-off between image visual quality and image fidelity. As shown in Fig.\ref{fig7}(a), using smaller values of $\tau_d$ gives higher fidelity (indicated by higher PSNR and SSIM) but worse visual quality (indicated by higher LPIPS scores). Conversely, a larger $\tau_d$ improves the visual quality but downgrades the fidelity. Fig.\ref{fig7}(b) shows the corresponding images. Smaller $\tau_{d}$ values, such as 0.0 and 0.2, produces blurry images with reduced tumor contrast compared to the ground truth. Setting $\tau_{d}$ between 0.8 and 1.0 achieves sharpness close to the ground truth. According to the curves in Fig.\ref{fig7}(a), increasing $\tau_{d}$ from 0.9 to 1.0 yields marginal enhancement in visual quality (LPIPS) but at a significant cost to fidelity (PSNR and SSIM). Therefore, we adopted $\tau_{d}$=0.9 for the entirety of this work.

\subsubsection{Image quality assessment from neuroradiologists}
Fig.\ref{fig_rad_exp1} shows neuroradiologists' ratings to demonstrate the clinical benefits of the super-resolution images given by FTDDM. We presented 2 experienced neuroradiologists with pairs of low-resolution and super-resolution images in a blind way (not labeled) from 20 slices, evenly selected from 10 patients. Each slice contains images of 7 metabolites, summing to a total evaluation of 140 image pairs. These images spanned five upscaling factors, $s$=8.0, 6.4, 5.3, 4.6 and 4.0, for a thorough evaluation. Conspicuity of each metabolite (tCr, tCh, NAA, Glu, Ins, Gln and Gly) was judged relative to the ground truth and scored as: 10 (poor), 30 (bad), 50 (fair), 70 (good) and 90 (excellent) \citep{chow2016correlation}. Using all metabolites, an overall sharpness and the overall conspicuities of gray matter, white matter, ventricles and lesion were judged relative to the ground truth and scored using the same scoring system. For the slices in which ventricles or lesion were not detected by the neuroradiologists, an ``NA" was given. We observe that the scores for the low-resolution images ranged between 10-70 for Neuroradiologist 1 and predominantly stood at 50 for Neuroradiologist 2. After being super-resolved by FTDDM, Neuroradiologist 1 rated over 87\% of the images as 90 (excellent), and Neuroradiologist 2 rated over 96\% of the images as 90, across all evaluation metrics. This reveals a significant enhancement compared to the low-resolution images. Fig.\ref{fig_allmets} shows the low-resolution, super-resolution and ground truth images of 7 metabolites from a slice containing a lesion, which is an example of what was presented to the neuroradiologists for image quality assessment. It is obvious that the lesion is less conspicuous in the 8$\times$8 zero-filled low-resolution images, and FTDDM successfully reconstructed the lesion contrast in tCr, tCho, Ins, Gln and Gly. This affirms why all super-resolution images were scored 90 (excellent) for lesion conspicuity. Moreover, the super-resolution images exhibit clear contrast between white matter and gray matter. This is especially observable in tCr, NAA and Glu maps, where concentrations are higher in gray matter. These metabolic characteristics are consistent with the ground truth images and the biological facts \citep{pouwels1998regional}.

\subsubsection{Determining the lowest input resolution}
Fig.\ref{fig_rad_exp2} presents neuroradiologists' ratings for images at three extremely low resolutions to determine the ``breakpoint" of our method. Low-resolution images at 2$\times$2, 4$\times$4 and 6$\times$6 predominantly scored 10 across all metrics, indicating very poor quality. Super-resolution results of the 6$\times$6 images mostly reach scores of 70 and 90, which is a significant enhancement. For 4$\times$4, the super-resolution images were scored 50 or 70, indicating borderline acceptability. However, when it comes to 2$\times$2, many super-resolution images were scored 50 or below. Supplementary Fig.S1 provides examples of these low-resolution and super-resolution images, highlighting that super-resolution images of 2$\times$2 do not align well with ground truth. Notably, tumor contrast begins to emerge in the super-resolution images of 6$\times$6. As a result, our confidence in the super-resolution results is retained only for the low-resolutions higher than 6$\times$6$\times$39  (equivalent to approximately 36$\times$36$\times$3.4mm$^3$); below this, our method might fail. For future applications, to achieve good super-resolution image quality, it is recommended that low-resolution MRSI data is acquired at resolutions finer than 36$\times$36$\times$3.4mm$^3$ (assuming similar acquisition techniques).

\subsubsection{Clinical values}
Lastly, it is worth noting that our 3D high-resolution MRSI datasets were acquired on a 7T MRI scanner using spatial-spectral encoding, which make it possible to acquire 3.4$\times$3.4$\times$3.4mm$^3$ resolution data in only 15 minutes. Using a standard phase encoding scheme on a clinical 3T MRI system (TR=450 ms and ellipsoidal k-space encoding) would require approximately 8 hours of scan time to achieve a similar resolution, which is clinically unfeasible. The proposed FTDDM, trained with our dataset, is able to transform lower resolution and faster scans into high-resolution metabolic maps. For example, an MRSI acquisition at 8$\times$8$\times$39 (or 27$\times$27$\times$3.4mm$^3$) can in theory be completed in approximately 2 minutes, which is a significant reduction in scan time from the 15 minutes requirement for a 64$\times$64$\times$39 (or 3.4$\times$3.4$\times$3.4mm$^3$) scan using our acquisition protocol. This reduction in scan time elevates MRSI's potential to be integrated in a routine clinical neuroimaging protocol. Furthermore, this time reduction also markedly decreases effects of motion, an important consideration when performing MRSI in clinical populations.

\section{Conclusion}
In this work, we introduce a novel Flow-based Truncated Denoising Diffusion Model (FTDDM) for super-resolution MRSI. Our method truncates the diffusion chain and hence accelerates the sampling process of the diffusion model. The noisy image at the truncation point is estimated directly from Gaussian noise via a normalizing flow-based network. The diffusion network is conditioned on the upscaling factor, facilitating multi-scale super-resolution. Additionally, a temperature parameter is incorporated into our diffusion model to allow sharpness adjustment. Experimental results on our self-developed in vivo \textsuperscript{1}H-MRSI dataset indicate that FTDDM outperforms other deep learning models and achieves over 9-fold acceleration compared to the conventional diffusion model without loss of image quality. Neuroradiologists' assessments confirm the excellent image quality given by the proposed method from clinical perspectives and assist in identifying the lowest operable resolution. To the best of our knowledge, we are the first to develop deep learning-based super-resolution models and conduct comprehensive evaluations on in vivo MRSI dataset. Future work to improve the generalization and robustness of the proposed method can involve the inclusion of more datasets acquired from other MRI systems \citep{nassirpour2018high,dong2024high} and/or from patients with different tumor types. The proposed model can also be extended to deuterium metabolic imaging (DMI) \citep{de2018deuterium,dong2020deep}.

\section*{CRediT authorship contribution statement}
\textbf{Siyuan Dong:} Conceptualization, Methodology, Software, Validation, Formal analysis, Investigation, Data curation, Visualization, Writing - original draft. \textbf{Zhuotong Cai:} Conceptualization, Writing - review \& editing. \textbf{Gilbert Hangel:} Data curation, Writing - review \& editing. \textbf{Wolfgang Bogner:} Data curation. \textbf{Georg Widhalm:} Data curation. \textbf{Yaqing Huang:} Writing - review \& editing. \textbf{Qinghao Liang:} Writing - review \& editing. \textbf{Chenyu You:} Writing - review \& editing. \textbf{Chathura Kumaragamage:} Writing - review \& editing. \textbf{Robert K. Fulbright:} Validation. \textbf{Amit Mahajan:} Validation. \textbf{Amin Karbasi:} Writing - review \& editing. \textbf{John A. Onofrey:} Conceptualization, Writing - review \& editing. \textbf{Robin A. de Graaf:} Conceptualization, Validation, Writing - review \& editing. \textbf{James S. Duncan:} Conceptualization, Writing - review \& editing, Supervision, Funding acquisition.

\section*{Data availability}
The code will be available online. Data will be made available on request.

\section*{Declaration of competing interest}
The authors declare the following financial interests/personal relationships which may be considered as potential competing interests: James S. Duncan is one of the Editors-in-Chief for Medical Image Analysis.

\section*{Declaration of Generative AI and AI-assisted technologies in the writing process}
During the preparation of this work the author(s) used GPT-4 from OpenAI in order to improve readability and language. After using this tool/service, the author(s) reviewed and edited the content as needed and take(s) full responsibility for the content of the publication.

\section*{Acknowledgments}
This work is part of Siyuan Dong's PhD research \citep{dong2024data}, supported by the National Institutes of Health under R01EB025840, R01CA206180 and R01NS035193. The data acquisition was supported by the Austrian Science Fund grants KLI 646, P 30701 and P 34198. 

\section*{Supplementary material}
Supplementary material was prepared and provided as a separate file along with the manuscript.

\bibliographystyle{model2-names.bst}\biboptions{authoryear}
\bibliography{refs}

\section*{Supplementary Material}

\renewcommand\thefigure{S\arabic{figure}} 
\setcounter{figure}{0} 

\begin{figure}[h]
\centering
\includegraphics[width=8.8cm]{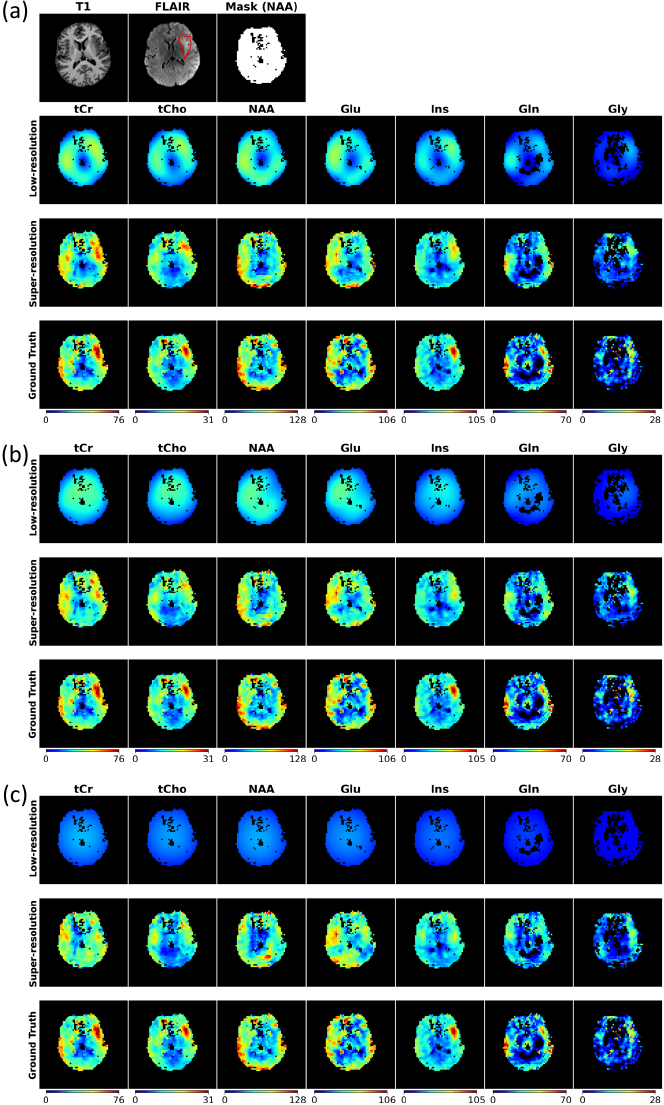}
\caption{Study of three extremely low resolutions. This is an example of the images provided to the neuroradiologists to give the scores in Fig.10. The low-resolutions at 6$\times$6, 4$\times$4 and 2$\times$2 are shown in (a), (b) and (c). The tumor is delineated by the red dashed line in FLAIR.}
\label{fig_S1}
\end{figure}

\end{document}